\documentclass{article}
\usepackage{graphicx}        
\begin{document}

\title{Linear Time Visualization and Search in Big Data using Pixellated Factor Space 
Mapping}

\maketitle

\author{Fionn Murtagh, University of Huddersfield, fmurtagh@acm.org}

\abstract{
It is demonstrated how linear computational time  and storage efficient
approaches can be adopted when analyzing very large data sets.  More 
importantly, interpretation is aided and furthermore, basic processing 
is easily supported.  Such basic processing can be the use of supplementary,
i.e.\ contextual, elements, or particular associations.  Furthermore 
pixellated grid cell
contents can be utilized as a basic form of imposed clustering.   For a given
resolution level, here related to an associated m-adic ($m$ here is a non-prime
integer) or p-adic ($p$ is prime) number system encoding, such pixellated mapping
results in partitioning.  The association of a range of m-adic and p-adic 
representations leads naturally to an imposed hierarchical clustering,
with partition levels corresponding to the m-adic-based and p-adic-based 
representations and displays.  In these clustering embedding and imposed cluster
structures, some analytical visualization and search applications are described.}

\section{Introduction}

While the ultimate aim of a great deal of data analytics is to have
clusters formed and studied, some open questions may need to be: to define
the dissimilarity or distance measure to use; then to define the 
cluster optimization criterion, or the hierarchical agglomerative 
clustering criterion.  This provides both motivation and justification
for the following.

Our approach here is to assume a factor or principal component space, 
thoroughly taking semantic relationships into account, and that is 
endowed with the Euclidean metric. For original data that is comprised
of categorical (qualitative) and quantitative values, Correspondence Analysis 
is most suitable.  Since the factor space is constructed through eigenvalue, 
eigenvector decomposition of the source data, it follows that if the number 
or rows, $n >> m$, the latter here being the number of columns, then 
the computational requirement is for $O(m^3)$ processing time.   This is 
likely to be achievable in practice. 

A particular benefit of Correspondence Analysis is its suitability for 
carrying out an orthonormal mapping, or scaling, of power law distributed
data.  Power law distributed data are found in many domains.  Correspondence
factor analysis provides a latent semantic or principal axes mapping.  Cf.\
\cite{ifcs2015}.

The case study used in this paper comes from the data studied in 
\cite{murtagh2016b}.  It is a large set of Twitter, social media, data. 
There are 880664 retained tweets, and 481 retained dates.  
Our major aims here are (1) to simplify, in an easily interpretable way,
the output, biplot or factor space planar plot, display; (2) to use an 
image-like approach to displaying such a plot; (3) to relate our data 
to forms of data encoding that are other than real-valued, and that are 
complementary to real-valued data.  In section \ref{marypadic}, a 
first stage of agglomerative hierarchical clustering is at issue, but 
this is contiguity- or adjacency-constrained.  While the latter terms are
relevant, it is better expressed as being sequence-constrained where that
constraint applies to what is to be clustered (to be seen later, this 
applies to row or column sets of grid cells).  

\begin{figure}
\begin{center}
\includegraphics[width=12cm]{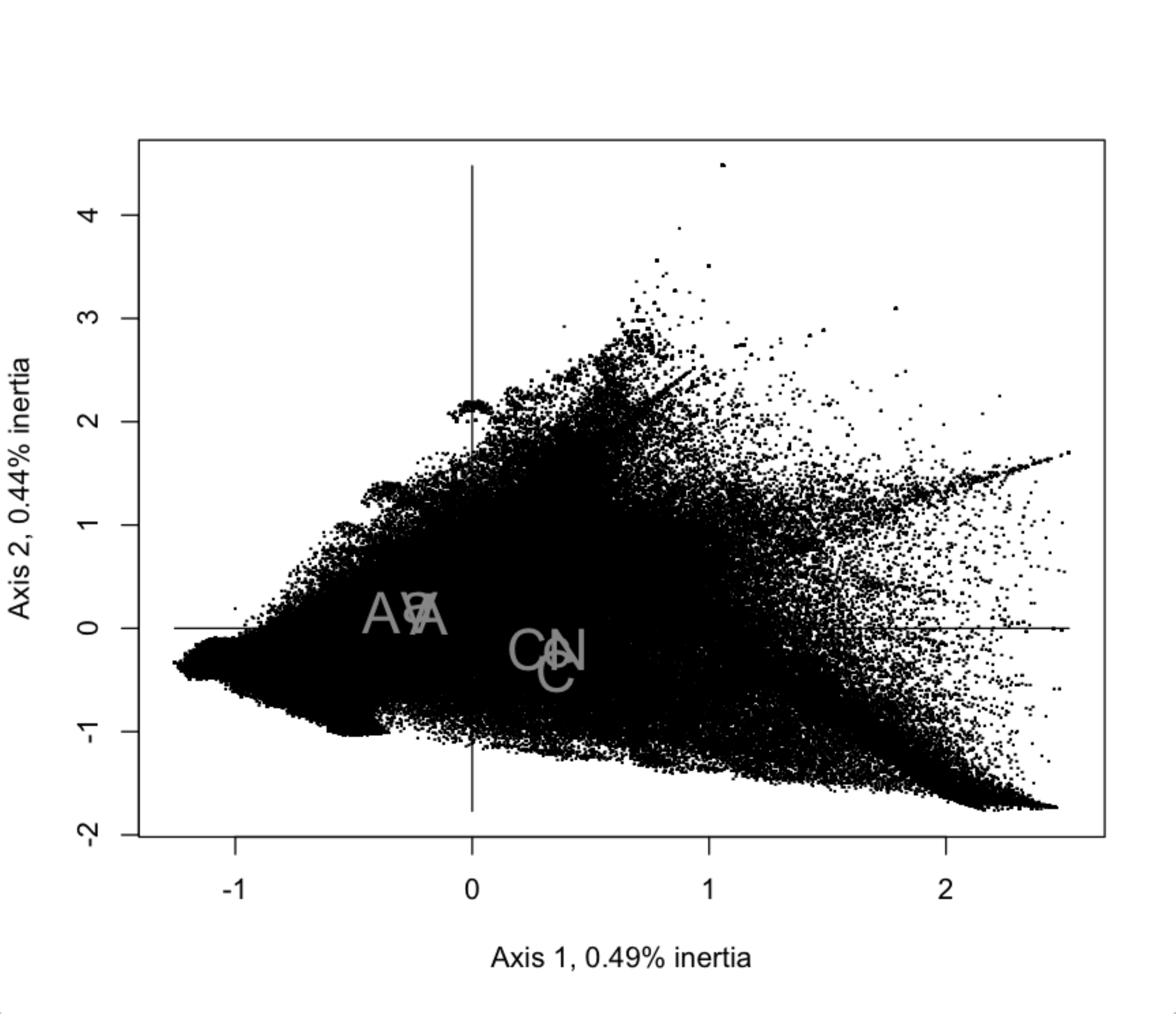}
\end{center}
\caption{880664 Twitter tweets projected on the principal factor, i.e.\
principal axis plane, with attributes projected.}
\label{Baire2Dfig1supp}
\end{figure}

Our priority is to consider here the two-dimensional, principal factor plane.  
This is produced from 880664 Twitter tweets, sometimes also termed micro-blogs.  
In Figure \ref{Baire2Dfig1supp}, there is this planar projection or mapping, that 
is also termed a biplot, with the mapping too of the three references in these 
tweets to the Cannes  film festival. Used here are ``C'', ``c'' and ``CN'' for 
different use of upper can lower case in these references to the Cannes film 
festival. Also in this mapping are references to the Avignon theatre festival, 
using labels ``A'', ``a'' and ``AV'', relating to the user of upper can 
lower case letters in the tweets.  

Our first task is to have an analogy of a two-dimensional histogram 
of what is displayed in the two previous figures.  This can be advanced
towards an image representation.  Pixellation of a spatial domain can
be displayed using heatmaps, or false colour coding, implying the 
predominant visualization role for such display. 

\section{Algorithm for Determining and Contextual Displaying of a 
Two-Dimensional Histogram}

Pixellating the data is equivalent to hashing, and for an image representation, 
the viewpoint is employed, 
of having this considered as a two-dimensional (2D) histogram.

\subsection{Pixellating the Mapped Data Point Cloud}

Firstly, given that the factor space is endowed with the Euclidean distance,
the coordinates of the projected cloud of points can be rescaled.  One 
reason for this is to have a standardized algorithmic processing approach.  
If there were to be particular topological patterns (e.g.\ the horseshoe effect,
curvilinear pattern related to cluster ordering through embeddedness) or other 
targeted cluster morphologies (shapes, whether in two dimensions or in the full
dimensional space), the the analytics would focus on such patterns.  Here, 
algorithmically, the general and generic objective is carried out by mapping 
the coordinates of the projected data cloud, in the factor plane, onto 
coordinates that are in the (closed lower bound, open upper bound) 
interval $[0, 1)$.   This mapping results from the interval of minimum value, 
maximum value, on each coordinate. 

This then allows pixellation of the rescaled, unit square area containing the 
mapped cloud of points.  This could be generalized also to unit volumes if
more than two factors are at issue.  Pixellation, i.e.\ imposing a regular
grid, can be displayed as a grid structure.  Furthermore, the pixel value is 
to be defined by the frequency of occurrence of points mapped into the given 
pixel area.  To be both precise about what is displayed and to be computationally
and storage efficient, the frequency of occurrence values, comprising our pixel 
values here, will be displayed.  

Such objectives in data analysis are largely consistent with the computational and 
interpretational advantages and benefits that are described in \cite{murtaghberry}.
First we may note that the heatmap display, provides colour coding of the data 
matrix values.  This is based on permutation of rows and columns with adjacency 
that is compatible with the hierarchical clustering of both rows and of columns.  
Such visualization of hierarchical clustering is at issue in \cite{zhangetal}.

Now, a similar view of analysis might start with a very large data array and decide
that a convenient and computationally efficient analysis process is as follows.  
Firstly, have the rows and columns permuted so that there is at least some relevance 
of the proximity taken into account by the permutation.  In \cite{murtaghberry}, 
this was based on principal factor projections. Then, secondly, the row and column
permuted input data array is considered such that all array values are the pixel values
of an image.  So the input data array is thus considered as an image.  Then 
computationally efficient processing can be undertaken on the image representation 
of the input data array.  Such processing is quite likely to involved a wavelet 
transform of the image, with filtering carried out in wavelet transform space.  This
provides a manner of determining clusters, and all is very relevant when visualization 
is an desirable outcome.  

In a sense therefore, at issue here, firstly,
is the carrying out of the factor space mapping, 
endowed with the Euclidean metric, and that can be computationally efficient if, 
for example, having $n$ rows, and $m$ columns, then eigen-reduction that determines
the factor space mapping, when $m << n$, is computationally, $O(m^3)$.  This can be quite
efficient, assuming that $m << n$.  Then all of the work here is, analogously, to map 
large data arrays into images, here directly based on the Euclidean metric endowed
factor space. 

Sample R code used for pixellation is available at this website, \\ 
{\tt http://www.multiresolutions.com/strule/papers}.

\section{Visualization through 2D Histogram Representation}

Following Figure \ref{Baire2Dfig1supp}, Figure \ref{Baire2Dfig5supp} shows how 
we can have displays that are both informative 
and also that provide alternative display capability for very large numbers of 
projection locations, i.e.\ mapped rows or columns, observations and attributes. 

In Figure \ref{Baire2Dfig5supp}, it seems visually that the high frequency
value in the grid cell that has a projection greater than 4 on axis 2, and
greater than 1 on axis 1, is unacceptably high in value.  This grid cell 
has 13333 projected points.  In fact we verified that there are 13333 overlapping
points here.  (Their values on factor 2 are all:  4.476693). 

A heatmap may be derived from this representation, that is grid-based and can
be characterized as a 2-dimensional histogram.   
The heatmap, being false colour coding of the data matrix being hierarchically
clustered, in regard to both the row set and the column set, is used in 
\cite{zhangetal}. 
Such a heatmap display
has a somewhat different objective, relative to what will now follow.
Our later aim is have the data structured to support search and retrieval.

\subsection{Varying Number Theoretic Representations of the Data: m-Ary and p-Adic 
Representations}
\label{marypadic}

Having pixellated the projected cloud of points, this is based on associating 
each projected point with its grid cell.   Integral to this is that the grid 
cell containing the projected point becomes an expression that labels each of 
its grid cell members.  This is analogous to projected point being a member of 
a cluster, and perhaps also it is analogous to the projected point being 
conceptually characterized by the superset of projected points associated with the
grid cell.  Since the grid cells are defined by default as decimal numbers, 
that will also be termed here, 10-ary, i.e.\ m-ary with m = 10, we will next 
considered alternative number theoretic representations.  With p being a 
prime number, and m being a non-prime integer, we will consider the best fitting 
grid cell mapping representations that are derived from the decimal or m = 10,
m-ary, representation.  We considered: m = 9, m = 8, p = 7, m = 6, p = 5, m = 4,
p = 3, p = 2, number representations.  

Considerable background discussion on p-adic number systems, and their role in various
domains, is in \cite{murtagh2017}.  Specifically using the algorithm now to be 
described, for closest fit of one number theoretic system to another, this is 
described in \cite{murtagh2016}.  

Algorithmically to move from an $m = 10$ $m$-ary representation to an $m = 9$ $m$-ary 
representation, we take the definition of the grid cells from their projections 
on factor 1 and on factor 2.  We first consider factor 1, i.e.\ axis 1, with 
identical reasoning applied to factor 2, i.e.\ axis 2.
Let $v$ be the constant interval between grid lines.
For $m = 10$, we have the axis 1 values as follows: $0, 0 + v, 0 + 2v, 0 + 3v, 
\dots, 0 + 9v$.   We now want to form grid cell boundaries for $m = 9$, so 
that on axis 1, we will have: $0, 0 + v^{\prime}, 0 + 2v^{\prime}, 0 + 3v^{\prime}, 
\dots , 0 + 0 + 8v^{\prime}$.  

Because we want a closest fit by the 9-ary representation to the 10-ary representation,
the former is based on the latter.  We take the least difference between the total 
sum of successive grid bins.  In effect, then, we merge these two successive grid bins.
By relabelling higher valued grid bin sequence numbers, this directly provides us with a
9-ary representation.  

The same approach is used for factor 2, i.e.\ axis 2.  Then we can proceed through 
further stages, to find a best fit 8-ary representation to the just determined
9-ary representation.  The following stage is to find a best fit 7-ary representation
to the just determined 8-ary representation.  This continues stagewise until we have
a 2-ary representation, i.e.\ a binary representation of the grid cell axis 1 and 
axis 2 projections.  

Further study of pixellation was for a $ 9 \times 9$ grid display, 
an $ 8 \times 8$ grid display, a $ 7 \times 7$ grid display, and so, continuing to
a $ 2 \times 2$ grid display.
The latter is associated with a binary encoding of the grid cell boundary coordinates. 

Thus, to state what is at issue here, it is visualization through 2D (two-dimensional)
histogram display, possibly accompanying m-adic and p-adic representation, in our 
mapped or represented data. 

We may wish to determine a rather good display of the supplementary elements 
relative to particular grid cells.  

\begin{figure}
\begin{center}
\includegraphics[width=12cm]{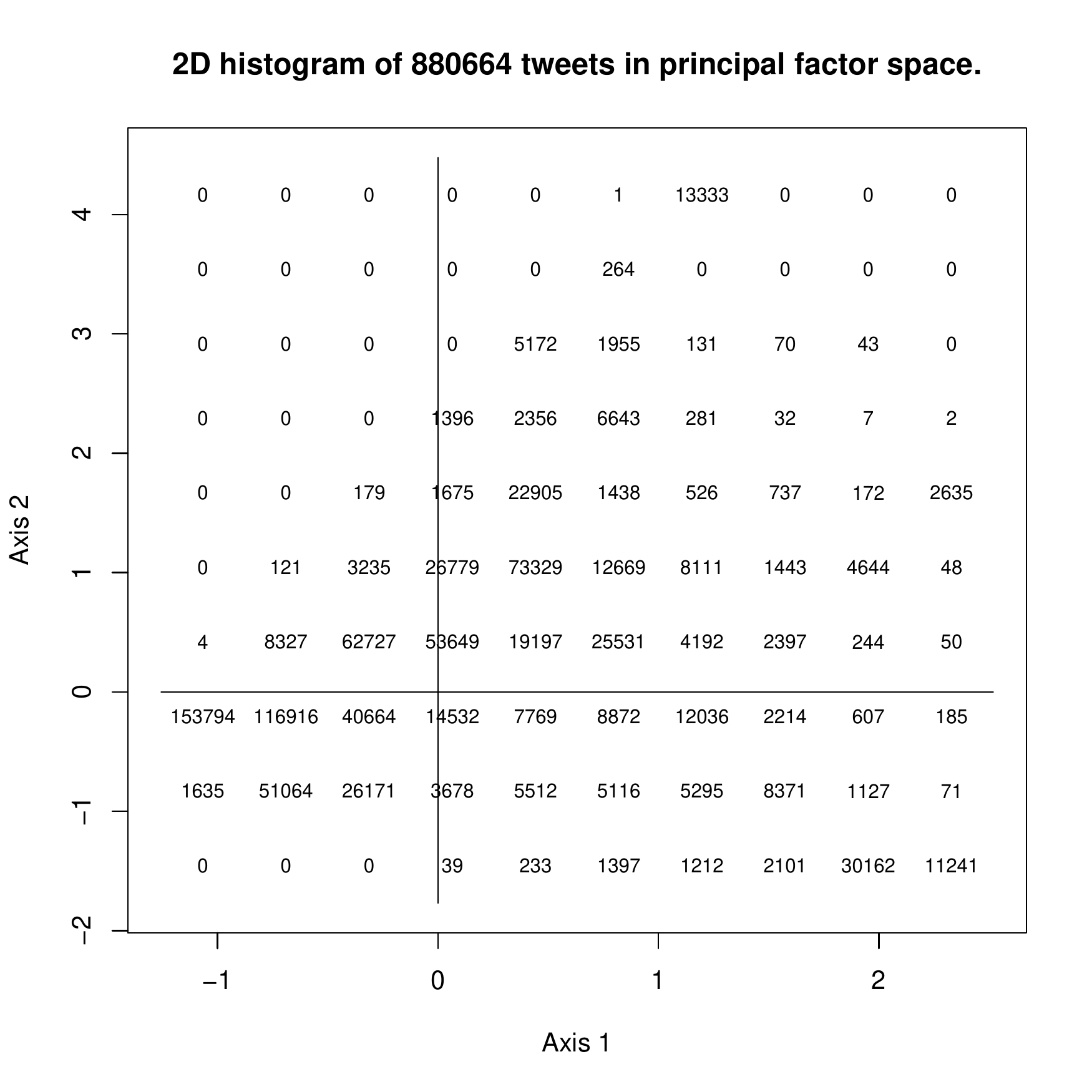}
\end{center}
\caption{The grid structure, used for the 2-dimensional histogram binning, is not 
displayed but the histogram bin values alone.}  
\label{Baire2Dfig5suppB}
\end{figure}

\begin{figure}
\begin{center}
\includegraphics[width=12cm]{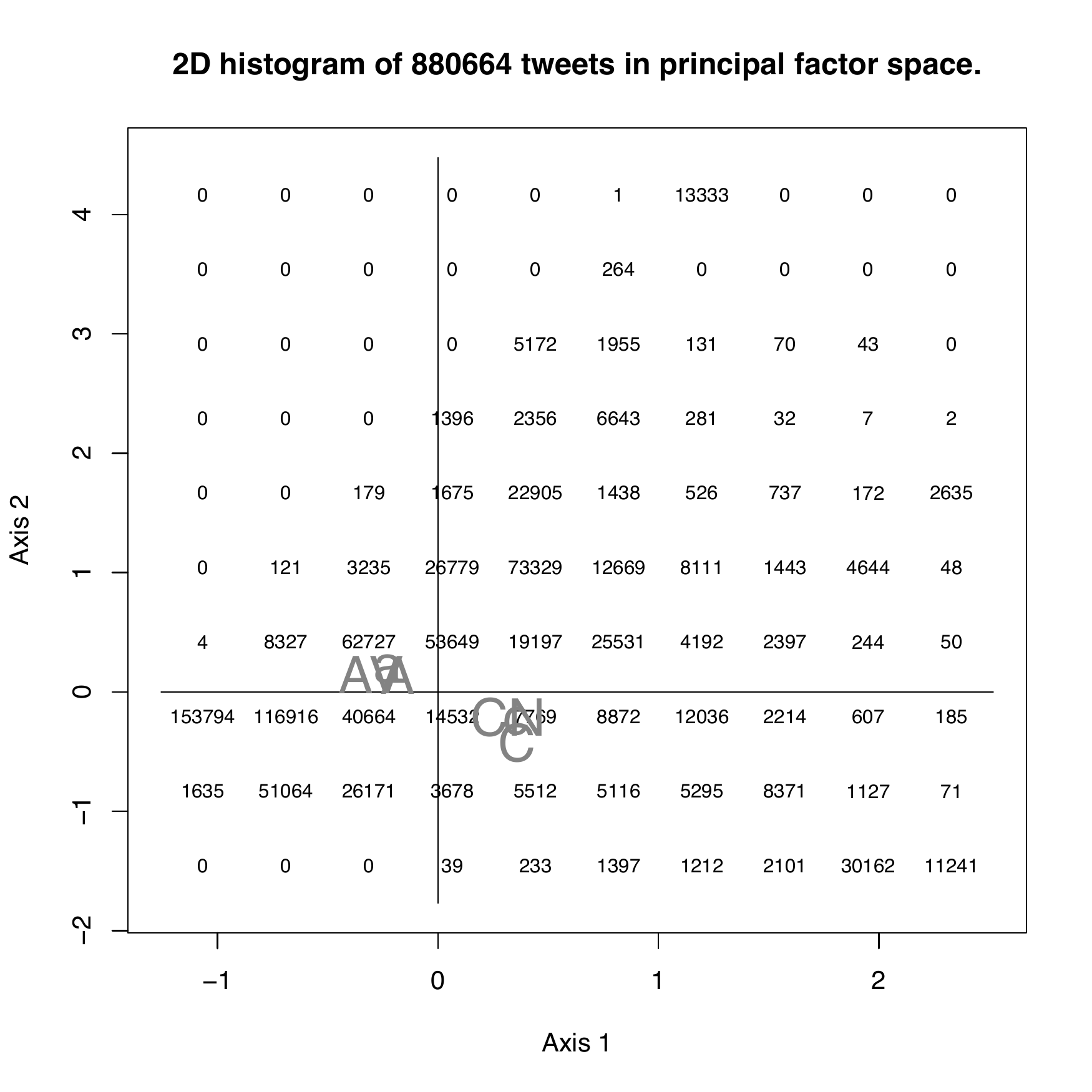}
\end{center}
\caption{As previous figure, Figure \ref{Baire2Dfig5suppB},  
with supplementary elements projected.
These projections, associated with tweet content, are related to the Cannes Film festival
and the Avignon Theatre festival.}
\label{Baire2Dfig5suppC}
\end{figure}

There can be the display of local densities using the projected elements.  
Taking previous outputs, Figure \ref{Baire2Dfig5suppB} displays local 
densities of the tweets, provided by the $10 \times 10$ grid.  The grid 
cells can be considered as three dimensional histogram bins.  Then in 
Figure \ref{Baire2Dfig5suppC}, the supplementary variables are projected,
relating to the words used for the Cannes Film Festival, and the Avignon
Theatre Festival.  At issue here is largely the display.  

\begin{figure}
\begin{center}
\includegraphics[width=12cm]{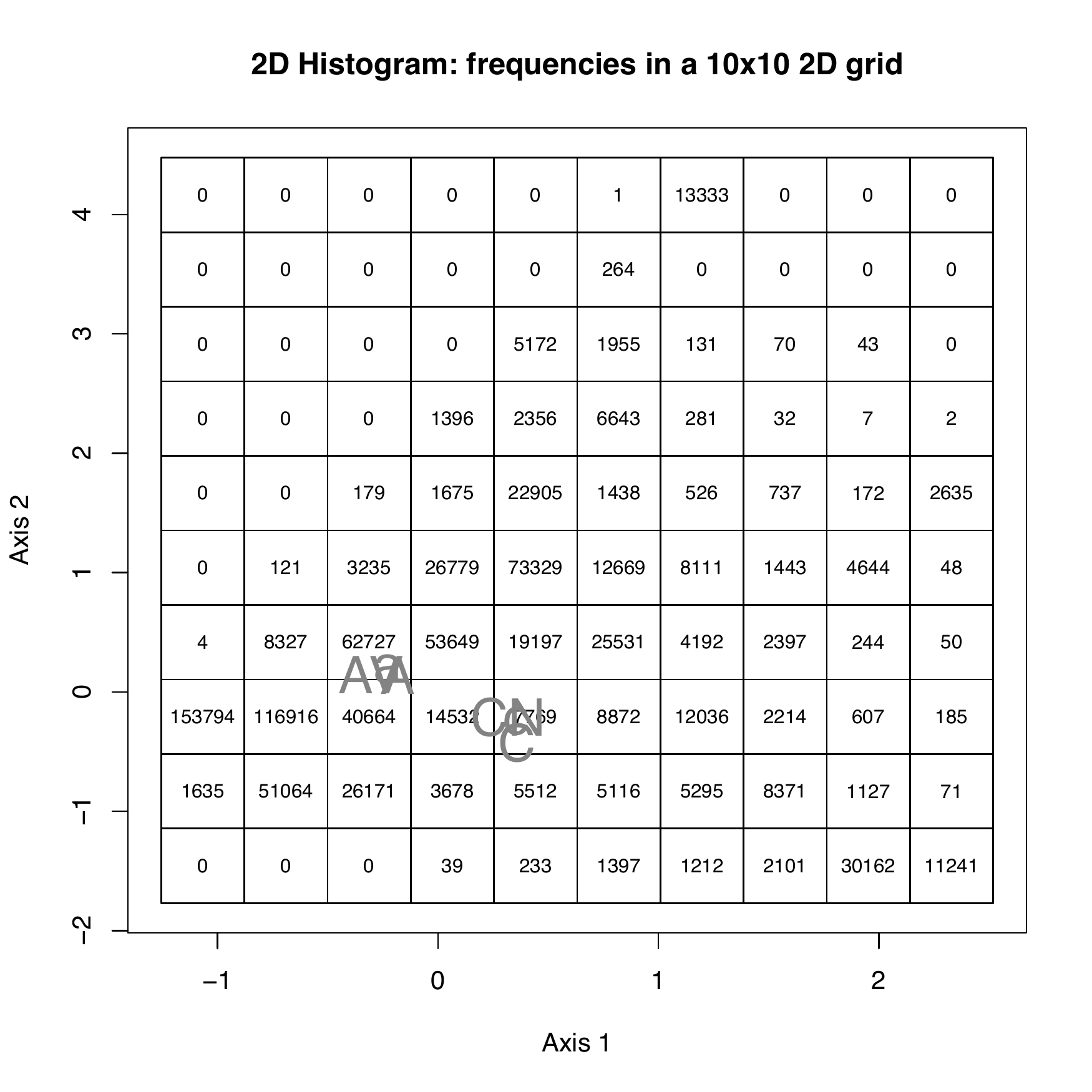}
\end{center}
\caption{A $ 10 \times 10$ grid display with supplementary elements projected.
These projections, associated with tweet content, are related to the Cannes Film festival
and the Avignon Theatre festival.}
\label{Baire2Dfig5supp}
\end{figure}

\section{Hashing and Binning for Nearest Neighbour Searching}

In \cite{heckmurtagh}, a framework for all that is at issue here was 
described with relevance in the fields of information retrieval and 
kindred areas related to search and retrieval.   For nearest neighbour
searching, what is discussed initially is: hashing, or simple binning or
bucketing.  The grid structure cell constitutes the way forward for 
searching in other neighbours of the query point that have also been 
mapped into the one grid cell. Some consideration may need to be given 
to one or more adjacent grid cells if the query point is closer to a
grid cell boundary, than it is to any potential nearest neighbour in the 
given grid cell.  With uniformly distributed data, here in the 2-dimensional
context, then it is noted how constant time, i.e.\ $O(1)$ 
time, is the expectation (statistical first order moment, mean) of the 
computational time.  A proof of this is in \cite{bentley}. 

It is to be noted, \cite{heckmurtagh}, p.\ 33, 
that when searching requires use of adjacent multidimensional grid cells, 
by design hypercube in their hypervolume, then this implies a computational 
limitation that increases with dimensionality.  In \cite{heckmurtagh}, 
for dimensionality reduction that supports nearest neighbour searching, 
reference is made to principal components analysis, non-metric 
multidimensional scaling, and self-organising feature maps.  In our
work here, Correspondence Analysis provides semantic mapping, i.e.\ 
the factor that is, in effect, the principal component space, with 
a unified scale for both observations and attributes.  

Extending this approach to higher dimensional spaces, there is the k-d 
(k-dimensional, for integer, k)
tree, or multidimensional binary search tree.  This is a balanced tree, 
by design, representation of the multidimensional point cloud.  
Stepwise binarization of the data is carried out using the median projection
on each axis.  However, the computational complexity of requiring the 
checking of adjacent clusters that are, by design, hyperrectangular, 
has the following consequence: computationally such an approach needs 
to be limited in the dimensionality.  Up to dimensionality of 8 is reported
on in the literature.  

Also discussed in \cite{heckmurtagh} are bounds on the nearest neighbour
distance, given a candidate nearest neighbour; other such bounding using 
metric properties, especially the triangular inequality; branch and bound
is the at issue in such methodology; for high dimensional,
sparse, binary data, and where binary here represents presence or absence 
values in, e.g., keyword-based document data (as an example, there could be 
10,000 documents, crossed by the presence or absence of 10,000 keywords).  
For the high dimensional, sparse, binary data, such data has always been 
traditional in information retrieval, and what is used is the mapping of 
documents to keywords, and the inverted file that is the mapping of keywords
to documents.  Cluster-based retrieval can extend some of these approaches. 
Further discussion in \cite{heckmurtagh} is for range searching using the 
quadtree, for 2-dimensional images, the octree, for 3-dimensional data cubes 
(just one example here is a 3-dimensional image volume), and a quadtree 
implementation on a sphere, for spherical data. The latter is relevant for 
remotely senses earth data, and for cosmological data.  Range searching 
involves moving beyond location-oriented search.  

\section{Multidimensional Baire Distance}

An open issue, motivated by this work, is to aim at having a multidimensional
Baire distance.  This could be based on the following.
Take a full factor space, perhaps with 5 factors retained (as it the default 
in the FactoMineR package in R), and for such labels here as C, A, etc., 
looking at grid binned factor pairwise (biplots) supplementary mapping.  This 
is to to see what grid cells are relevant 
for the supplementary elements.  But, based on this approach, we may have 
supplementary rows or individuals, that with Twitter data, is to then do 
digit-wise mapping of tweets against the selected supplementary elements.

In general, related to such a multidimensional Baire distance is the Baire
distance formulated for multi-channel data, i.e.\ hyperspectral images, and 
used for machine learning (Support Vector Machine, supervised classification)
in \cite{bradley}.

\section{Conclusion}

A central theme of this work can be expressed as follows: performing data 
mapping that results in a domain-relevant data encoding.  One aim of the work 
here has been to benefit from just how very evident it is, that human 
visual-based information, and both measure and approximated data, become very 
efficient as well as effective.  At issue are: image, display, biplot.
Further practical benefits are 
demonstrated in \cite{murtaghberry}, by representing the data to be analyzed
as an image, and thereby carrying out wavelet transform based filtering, and object
detection, and so on.

Informally expressed, therefore, we may state that this work is in relation 
to the visualization of data, that accompanies having the data verbalized: 
see \cite{blasius} for this phrasing. The application objectives cover data 
mining (as contrasted with supervised classification and mainstream data 
mining), data analytics (encompassing both processing and output), and 
inductive reasoning (what the analysis achieves).  Furthermore, the computational
complexity of the processing and all of the implementation is implicit in this 
work.

Largely, the terms used here, pixellation and a 2D histogram, are identical.
Search in Big Data is the basis for matching, such as in nearest neighbour 
matching, and associated or relevant data querying.  Although not always 
done, from a mathematical sciences point of view, well-based, past work
ought to be cited.  In \cite{murtagh1985}, chapter 2 entitled ``Fast nearest
neighbour searching'' covers multidimensional binary search tree structuring 
of the data, and hashing for nearest neighbour or best match searching, 
accompanied by implementation optimization, and with twenty two references 
on these, and directly related, themes.   

All that is at issue here is open to the possibility of implementation using
distributed computing.  While further research will deal with case studies 
arising and motivated by this work, a further aim will be to have distributed
computing implementations also set up and functioning operationally.

\end{document}